# Resource-Aware Replication on Heterogeneous Multicores: Challenges and Opportunities

Björn Döbel, Robert Muschner, Hermann Härtig
*Technische Universität Dresden*
*Dresden, Germany*
{*doebel,robemusc,haertig*}*@tudos.org*

## Abstract

Decreasing hardware feature sizes and increasing heterogeneity in multicore hardware require software that can adapt to these platforms' properties. We implemented ROMAIN, an OS service providing redundant multithreading on top of the FIASCO.OC microkernel to address the increasing unreliability of hardware. In this paper we review challenges and opportunities for ROMAIN to adapt to such multicore platforms in order to decrease execution overhead, resource requirements, and vulnerability against faults.

## 1 Introduction

Commercial-off-the-shelf (COTS) hardware components are becoming more powerful and complex with every hardware generation. Decreasing transistor sizes are an enabler for these developments, because vendors can now add ever more functional units with lower energy consumption. The downside of this development is that processors become more vulnerable to permanent and transient hardware errors [4]. This trend is expected to increase and poses a serious threat to future hardware generations [10].

While hardware fault-tolerance mechanisms to mitigate these reliability issues exist [3, 8], COTS vendors try to avoid such extensions because they make hardware more expensive to produce. Researchers addressed this problem by proposing software-level fault tolerance methods that run on COTS hardware. These methods come in the form of compiler extensions [15] and extensions to the runtime environment [18].

The increasing availability of varying processor units leads to a second trend: today's hardware platforms are becoming more heterogeneous. Non-uniform memory architectures and cache hierarchies require new forms of scheduling threads and data [22]. The availability of specialized compute units and general-purpose GPUs increases the complexity of deciding when and where to run an application workload [13].

We designed ASTEROID, a fault-tolerant operating system architecture as shown in Figure 1. This architecture combines the FIASCO.OC microkernel, the corresponding L4Re user-level runtime environment, and ROMAIN [6], an operating system service that provides replicated execution for binary applications using software-implemented redundant multithreading [14].

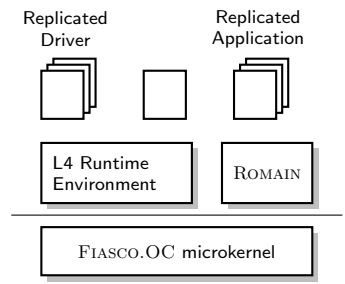

Figure 1: ASTEROID Resilient OS Architecture

In this paper we survey ROMAIN in Section 2. Thereafter we discuss three challenges with respect to adapting to heterogeneous hardware and evaluate how ROMAIN may aid or benfit from these scenarios. In Section 3 we show how the underlying cache hierarchy interacts with replication performance. Thereafter we present a method to adapt the number of ROMAIN replicas to environmental vulnerability conditions in Section 4. Last, we show how the ROMAIN software architecture can be mapped to CPU cores with different resilience levels in order to better protect ROMAIN's Reliable Computing Base (RCB) in Section 5.

## 2 Replication as an OS Service

Compiler extensions that provide software-implemented fault tolerance require the protected software's source code to be available for recompilation. This is often impossible as many vendors distribute software as binaries. Replication-based fault tolerance schemes do not share this requirement and can therefore protect binary applications.

We implemented ROMAIN, an extension to FIASCO.OC's L4 Runtime Environment (L4Re).[1] ROMAIN provides a software implementation of redundant multithreading [14]. As shown in Figure 2, multiple replicas of a protected application run in isolated address spaces on different CPU cores in a multicore system. A master process controls these replicas by managing their resources and intercepting all their communication to the outside world. The master thereby makes sure that replicas always obtain identical inputs and produce the same outputs unless they are affected by a hardware fault.

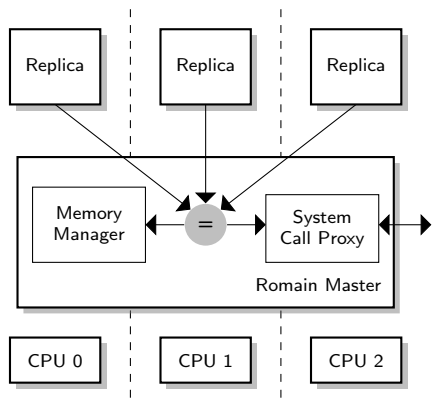

Figure 2: ROMAIN Replication Architecture

ROMAIN executes replicas using FIASCO.OC's virtual CPU (vCPU) mechanism [11]. This OS feature allows the master process to intercept any externalization event (e.g., system calls, page faults, CPU exceptions) that is generated by a replica. Replicas therefore never directly interact with applications outside their *sphere of replication* [14].

When encountering an externalization event, the master process first compares the replicas' states to validate that they still match. After successful validation, the replicas' externalization event is handled. For OS system calls, handling means that the master performs the respective system call on behalf of the replicas, while the replicas remain blocked. Once the system call returns, the master adjusts the replicas' states as if they had issued the system call themselves. Thereafter, the replicas resume independent execution until they reach their next externalization event.

Microkernels — such as FIASCO.OC — perform all memory management in user-space applications [2]. ROMAIN leverages this property to maintain full control over the resources owned by each replica. Any replica request to create a new kernel object is intercepted and emulated by the master process. Furthermore, the master acts as the replicas' memory manager. It keeps track of which regions in virtual memory are attached to the replicas and services any page faults that arise while replicas execute.

ROMAIN distributes replicas across the available CPU cores in a multicore system. This has two advantages: first, this approach adds another layer of redundancy to the system. If one of the CPU cores encounters a permanent hardware fault, only a single replica is affected by this problem. ROMAIN can detect the resulting error and correct it for instance by reassigning the faulty replica to another CPU core. Second, using multiple CPUs allows replicas to execute concurrently as long as they only perform internal computations. This approach therefore also minimizes the runtime overhead for replicated execution.

Our initial implementation of replica assignment distributed replica threads sequentially across the available CPUs, starting at CPU 0. We will see in the following section that this strategy – which does not incorporate any knowledge about the underlying platform and its properties – is far from ideal.

## 3 Adapting to Resource Requirements

In modern multicore platforms, CPUs are often distributed across multiple sockets. CPUs on a socket share caches and local memory. Inter-processor interrupts (IPIs) between CPUs on the same sockets are delivered faster than to CPUs on a different socket. To optimize runtime overheads, ROMAIN needs to be aware of the cache and communication hierarchy when placing replicas on CPUs.

We performed an experiment on a multicore system containing two sockets, each containing six Intel Xeon X5650 CPUs clocked at 2.66 GHz. We analyzed the sources of replication overhead for ROMAIN and found IPIs that are sent between replicas and the master for every externalization event to be large contributors of overhead. For the platform in question we found that IPIs require an average of 5,900 CPU cycles for intra-socket communication and 14,300 CPU cycles for inter-socket communication.

Based on this observation we implemented a core assignment algorithm in ROMAIN that tries to place replicas on CPUs on the same socket in order to save IPI overhead. With these optimizations in place, we executed 11 of the 12 SPEC INT 2006 benchmarks in ROMAIN.[2] We show three classes of results in Figure 3 and compare them to native execution (i.e., without ROMAIN). First we executed a single instance of each benchmark using ROMAIN. This setup does not provide any fault tolerance, but solely measures the runtime overhead induced by intercepting system calls and proxying them through the ROMAIN master. The second class of experiment shows the overhead for running two replicas (DMR), providing error detection capabilities. Finally, a third set of experiments shows ROMAIN running in triple-modular-redundant (TMR) mode. This last setup shows the required overhead for achieving fault tolerance against a single faulty replica.

---

[1] http://l4re.org

[2] We left out 483.xalancbmk because it uses deprecated C++ STL features that are not supported by L4Re's standard C++ library.

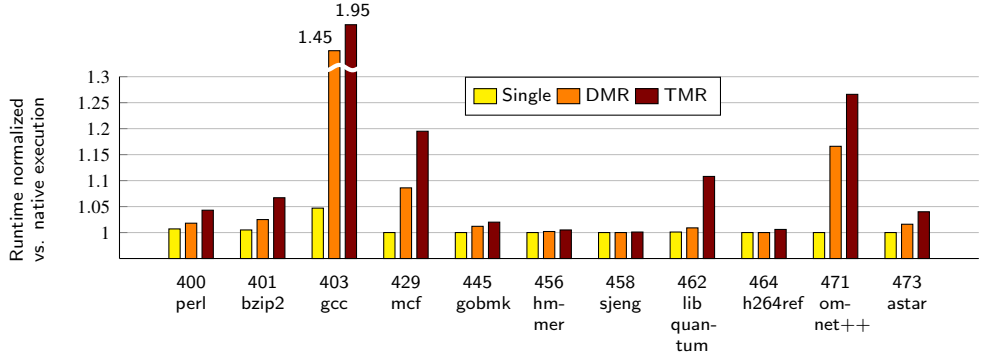

Figure 3: Overhead for replicating the SPEC INT 2006 benchmarks with one, two, and three replicas compared to native execution. Geometric mean overheads: $GM(DMR) = 0.66\%$, $GM(TMR) = 2.51\%$

Replication with ROMAIN is acceptably cheap in most cases – the geometric mean overhead for TMR execution is at 2.51%. However, the results show that four of the benchmarks induce considerably higher overheads and we therefore investigated these benchmarks more thoroughly. The high overheads for the 403.gcc benchmark can be attributed to its exceptional memory access patterns, which cause high memory management overheads as well as trigger errors in the current ROMAIN prototype's memory management.[3]

We inspected the 429.mcf, 462.libquantum, and 471.omnet++ benchmarks more closely using hardware performance counters. We found these benchmarks to cause a huge amount of last-level cache misses. All replicas share a single L3 cache when placed on the same CPU socket and this cache then becomes a replication bottleneck. We adapted ROMAIN's CPU placement algorithm to distribute replicas across our two CPU sockets, thereby giving replicas access to two separate L3 caches.

We then re-ran the benchmarks and show the improved benchmark results in Figure 4. The experiment confirms that our initial CPU assignment strategy hurts cache-bound applications and that these applications should be distributed across CPU sockets.

**Adaptation Challenge:** Assigning replicas to CPUs in a heterogeneous platform can significantly influence replication overhead. Our current strategy assigns replicas to CPUs statically. Hence, ROMAIN needs to rely on user knowledge to configure the proper assignment. Our vision for future versions of ROMAIN is to monitor cache miss rates of replicas at runtime. The master process can thereby distinguish between communication-bound and cache-bound applications. The former benefit from being placed on a single socket, whereas the latter benefit from being distributed over all sockets to increase cache availability.

Reliability adds an orthogonal perspective: Replicas running on the same socket may be affected by faults that hit the whole socket. Therefore distributing replicas across sockets may increase tolerance against these faults. On the other hand, if the communication connection between two sockets fails, distributed replicas may no longer synchronize whereas replicas on the same socket still function correctly. Such interactions between the hardware fault model and replica placement need to be further investigated.

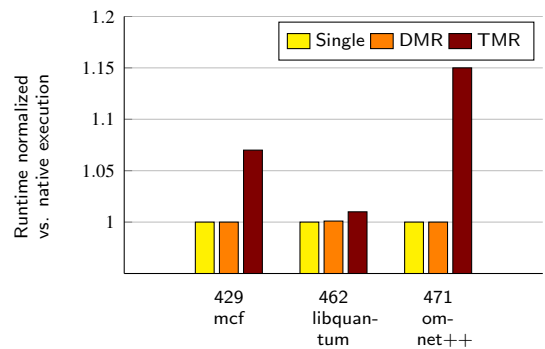

Figure 4: Overhead for replicating cache-bound SPEC INT 2006 benchmarks when distributing replicas across different CPU sockets

## 4 Adapting Resource Consumption

While the mechanism introduced in the previous section allows us to optimize replication overhead, it does not solve a more general problem of replicated execution: executing $N$ replicas requires roughly $N$ times the amount of resources of single-instance execution. Even though modern multicore platforms provide plenty of processor and memory resources, we would like to reduce resource consumption for replication in order to save energy.

---
[3] We nevertheless show these numbers for completeness.

Statically replicated applications need to configure the number of replicas (N := $2 \times num_{faults} + 1$ [17]) for the worst possible case. However, in many situations the protected system will not suffer from worst-case conditions all the time. An embedded device might for instance be designed to work in high-temperature or high-radiation conditions, but live in non-hazardous environments for most of its execution time. In this case we would like to switch on replication only during periods of hazardous operation and save replication resources at all other times.

From the perspective of a software developer, different parts of a program may have different criticality and vulnerability levels. It may therefore be useful to adapt the number of running replicas depending on the criticality of code that is executed at the moment. Program annotations or program vulnerability analysis [20] may give the replication system hints about when and how to adapt.

Based on these observations we assume that an external observer (e.g., a sensor determining hazardous situations) or the application itself (using specific compiler-inserted vulnerability hints) is able to notify the ROMAIN master process about the need to adapt replica counts. We added mechanisms to ROMAIN that interpret this information and adapt replica execution accordingly.

In order to decrease the number of replicas, ROMAIN waits for the next replica externalization event. At this point we know all replicas have reached identical state, otherwise ROMAIN would trigger error recovery. This means that the replicas have the same instruction pointer, register and memory content. Furthermore, no hardware I/O is in flight.

ROMAIN then puts one replica to sleep, releases all memory consumed by this replica, and resumes execution with $N - 1$ replicas. To increase replica count, ROMAIN wakes up a sleeping replica when getting a respective hint. The newly woken replica's CPU state is set to the state of all other replicas. New memory regions are allocated and their content copied over from a running replica. Then, ROMAIN resumes execution using $N + 1$ replicas.

ROMAIN's mechanism for dynamically adjusting replicas allows us to adapt resource and energy consumption to the actual reliability needs of a replicated application. However, we found that releasing and reallocating memory regions may cause significant overhead because of the additional memory management operations required for every replica modification. For the case of decreasing replicas, we hide the latency for freeing memory by performing these duties in a background worker thread that runs concurrently to the other replicas.

For the case of increasing the number of replicas we added a copy-on-write (COW) mechanism to ROMAIN. Instead of allocating new memory regions during replica startup, we make already existing regions available to the respective replica and mark them read-only. Only if a replica writes to this region we perform an actual reallocation and copy data. This mechanism reduces replica startup overhead, but has implications to ROMAIN's hardware requirements.

Our initial implementation of ROMAIN replicates all memory resources allocated by a replicated application. This approach allows replicas to execute without the need for synchronizing upon every memory access and therefore decreases replication overhead. From a reliability perspective, we can furthermore refrain from requiring costly ECC-protected memory, because replicated memory resources allow detection and correction of memory errors as well. In the case of COW memory regions, this guarantee no longer holds. The number of replicated memory regions may be smaller than the number of available replicas. A memory fault in a COW region may be seen by multiple replicas and can therefore lead to a situation where the majority of replicas makes their decisions based an this erroneous value.

**Adaptation Challenge:** ROMAIN provides mechanisms to dynamically adapt the number of replicas at runtime. Fast memory management using a COW mechanism only works if ROMAIN runs on a hardware platform providing ECC-protected memory. Otherwise we have to fall back to allocating and copying memory immediately during replica wakeup. System designers therefore need to address another challenge: Are they willing to get dynamic replication at the cost of increased overhead for replica creation? Or would they prefer to trade this overhead for increased production cost and energy consumption when using memory ECC?

## 5 Adapting to Hardware Vulnerability

The ASTEROID architecture uses replication to protect user applications against the effects of hardware faults. However, a subset of the system — consisting of the OS kernel and the ROMAIN replication servcie — is not covered by these protection mechanisms. This is problematic for the whole system's reliability: After all, these components are required to execute correctly in order to properly implement replication. We call these components the *Reliable Computing Base (RCB)* [7].

We believe that heterogeneous hardware can aid in protecting the RCB. Today's multicore platforms use heterogeneity to provide dedicated compute cores (such as general-purpose GPUs [16]) or energy-efficient processor alternatives (such as ARM's big.LITTLE [1]). The Error-Resilient System Architecture furthermore pro-processor messages by having the helper thread poll a shared-memory region for new data (e.g., using x86' `monitor/mwait` set of instructions).

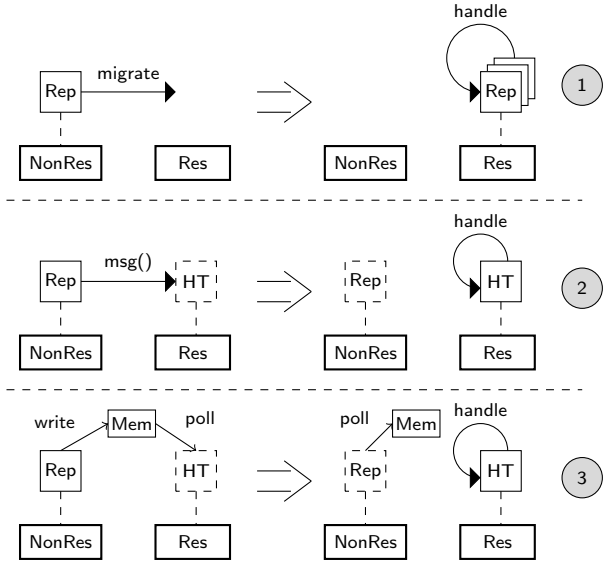

Figure 5: Notification variants: 1) Replica migration 2) Asynchronous Notifications 3) Shared-memory polling

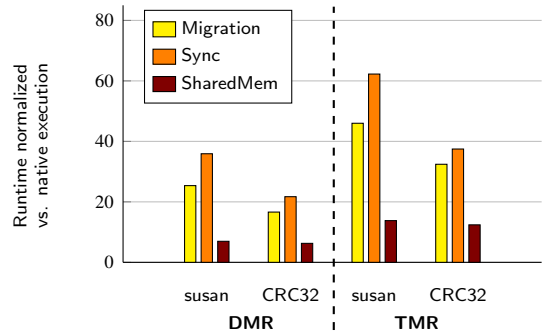

Figure 6: Runtime overhead for two high-overhead benchmarks from the MiBench benchmark suite when run with different inter-replica notification methods

posed multicore chips that comprise CPUs with different reliability levels [12]. We therefore assume that future heterogeneous platforms will provide resilient (*ResCore*) and non-resilient (*NonResCore*) CPUs on the same chip.

We propose to split the ASTEROID software into a non-resilient and the RCB layer [5]. The non-resilient layer is protected using the replication mechanisms described in the previous sections. The RCB layer is protected by running the respective software components on specific ResCores. The idea of running OS services on dedicated cores is nothing new and has previously been introduced by FlexSC [19] and fos [21]. Our contribution here is to use this split for protecting the RCB.

In Section 2 we explained that ROMAIN runs replicas concurrently on different physical CPU cores. To protect the RCB we have to distinguish between replica execution and master execution in ROMAIN. Replica code is scheduled by the ROMAIN master to run on any NonResCore. ROMAIN master code however needs to run on a ResCore. To implement replication, we now need an additional mechanism to transfer replica state between the resilient and non-resilient worlds.

We implemented three alternative mechanisms for state transfer as shown in Figure 5. Our first alternative uses FIASCO.OC's mechanism to migrate threads between CPU cores to migrate a replica to a ResCore for handling externalization events. The second alternative avoids migration cost and instead uses a RCB helper thread for every replica. Then, instead of migrating the replica, ROMAIN simply sends the replica state to the ResCore using a synchronous inter-processor message. Our third alternative avoids the cost of sending inter-processor messages by having the helper thread poll a shared-memory region for new data (e.g., using x86' `monitor/mwait` set of instructions).

Using these three mechanisms we were able to adapt ROMAIN to run RCB code and replica code on cores with different resilience levels. We evaluated the alternatives using benchmarks from the MiBench embedded benchmark suite [9]. We ran the benchmarks once using ROMAIN and then picked the two benchmarks with the highest replication overhead (susan, CRC32). For these two benchmarks we performed an experiment on a multicore machine assuming that CPU0 of this machine was a ResCore and all other cores were NonResCores. We then ran these benchmarks in ROMAIN using DMR and TMR setups. Figure 6 compares the overheads induced by the different notification mechanisms.

While our results show that shared-memory polling is the most efficient notification mechanism, this approach relies on correctly functioning shared memory between the non-resilient and resilient worlds. This may be another reliability issue because unreliable software might use such shared memory regions to overwrite data RCB components rely on. Hence, practical implementations may forbid shared memory to isolate protected data from unrprotected software. In this case we have to resort to migration or synchronous messaging.

**Adaptation Challenge:** ROMAIN allows running RCB code on resilient cores while replica code runs on non-resilient CPUs. Heterogeneous platforms may provide these hardware features. System designers then need to decide at which ratio to provide ResCores and NonResCores. Furthermore, hardware reliability mechanisms may be necessary to protect transmission of replica data to RCB software components.

## 6  Conclusion

In this paper we reviewed the ROMAIN OS replication service with respect to its adaptation capabilities for future heterogeneous multicore platforms. We showed that the heterogeneity introduced by such platforms leads to new optimization criteria when assigning replicas to physcial CPUs. We determined challenges that system developers face when applying replication on these platforms and showed opportunities that arise for improving overall system reliability by explicitly leveraging the platform's reliability properties.

## Acknowledgments

This work was supported by the German Research Foundation (DFG) as part of the priority program "Dependable Embedded Systems" (SPP 1500 - spp1500.itec.kit.edu).